\documentstyle[prl,aps,epsfig,floats,twocolumn]{revtex}
\def\bs{\vec{s}}

\def\bS{\vec{S}}
\def\bzeta{\vec{\zeta}} 
\def\boeta{\vec{\eta}}

\begin{document}
\draft


\title{Strange dynamics of domain walls and periodic stripes \\
along classical antiferromagnetic chains}
\author{Raphael Blumenfeld}

\address{Cavendish Laboratory, Madingley Road, Cambridge CB3 0HE, UK} 
\address{Email: rbb11@phy.cam.ac.uk} 
\maketitle 
\date{\today} 
\maketitle 

\begin{abstract} 
This paper addressses the kinetics and dynamics of a family of  domain wall 
solutions along classical antiferromagnetic Heisenberg spin chains at low energies. 
The equation of motion is derived and found to have long range position- and 
velocity-dependent two-body forces. 
A 'quiescent' regime is identified where the forces between walls are all repelling.
Outside this regime some of the interactions are attractive, giving rise to wall 
collisions whereupon the colliding walls annihilate.
The momentum of the system is found to be conserved in the quiescent regime and to 
suffer discontinuous jumps upon annihilation.
The dynamics are illustrated by an exact solution for a double wall system and a 
numerical solution for a many-wall system. 

On circular chains the equations support stable periodic stripes that can rotate as 
a rigid body. It is found that the stripes are more stable the faster they rotate.
The periodic structure can be destabilised by perturbing the walls' angular velocity 
in which case there is a transition to another periodic structure, possibly via 
a cascade of annihilation events.
 
\end{abstract}
\pacs{75.60.Ch}
\narrowtext

Spin chains with antiferromagnetic Heisenberg exchange coupling have attracted much 
attention in recent years due to the remarkable progress in synthesis methods, 
new experimental techniques, and a variety of possible applications. 
These chains exhibit a rich behaviour, much of which is not fully understood, either 
in the quantum or the classical regimes \cite{Botti}. 
Of particular interest is the low-energy regime which is amenable to some analytical 
treatment. It has been known for some time that this regime accommodates spin waves and 
instanton solutions \cite{SpinWavesInstantons} \cite{BPE}. 
Recently, however, it has been shown to also support a family of multitwist solutions 
\cite{BB} \cite{BBcurve} that extend a single twist found earlier \cite{BBD}. 
These solutions represent an arbitrarily number of domain walls that can travel along 
the chain. The multitwists are found by starting from the discrete Hamiltonian for the 
exchange coupling between nearest neighbour spins on the antiferromagnetic chain 

\begin{equation} 
H = \sum_i J \bS_i\bullet\bS_{i+1} 
\label{eq:Oi}
\end{equation}
and distinguishing between the even, $\bS_e$, and odd, $\bS_o$, sublattices. Defining the total, 
$\bzeta \propto \bS_e + \bS_o$, and staggered, $\boeta \propto \bS_e - \bS_o$ magnetisations  
(properly normalised) and taking the continuum limit, leads in the low energy regime to 

\begin{equation} 
{{\partial\boeta}\over{\partial t}} = \boeta \times {{\partial\boeta}\over{\partial x}}
\label{eq:Oii}
\end{equation} 
where $t$ is renormalised time and $x$ is the arclength parameter along the chain.
This equation translates into two coupled equations for the polar and azimuthal 
angles, $\theta$ and $\phi$, of the unit vector $\boeta$ which, in turn, are identical to 
the so called Belavin-Polyakov equations \cite{BPE} in the context of the nonlinear 
sigma model. The resemblance is not complete because in this context the equations are 
derived via the Bogomol'nyi derivation \cite{Bogo} and consequently come complete with 
uniform boundary conditions. Not do in the context of the antiferromagnetic chain where 
the boundary conditions can be arbitrary.

Introducing $\cos\theta = {\rm tanh}\psi$, whereupon 
$\boeta = (\cos\phi,\sin\phi,{\rm sinh}\psi)/{\rm cosh}\psi$ \cite{BB}, the coupled equations 
are transformed into 

\begin{equation}
{{\partial\psi}\over{\partial x}} = {{\partial\phi}\over{\partial t}} \ \ \ ; \ \ \
{{\partial\psi}\over{\partial t}} = - {{\partial\phi}\over{\partial x}}
\label{eq:Oiv}
\end{equation}
In the absence of sources, which give rise to logarithmic terms (corresponding to n-instanton 
solutions \cite{BPE} \cite{BBcurve}), the solutions are all the analytic functions in 
the $x-t$ plane. 
The smooth behaviour of these functions, once operated on with the hyperbolic function, yields 
multitwists in the angle $\theta$ along the chain where the nodes of $\psi$ are. 
These multitwists correspond to domain walls in the staggered magnetisation $\boeta$.

The aim of this paper is to analyse the dynamics of the domain walls corresponding to these twists.
Specifically, I will derive the equation of motion of the walls, showing that they interact via long 
range terms that are position and velocity dependent. It will be argued that there is a 'quiescent' 
regime where the system of walls conserves its global momentum. 
When wall velocities become sufficiently high the system exits this regime and domain walls 
develop mutual attarction, leading to potential annihilation events and consequently to rich 
dynamics. Upon annihilation, momentum conservation is broken for the entire system but remains 
intact for the remaining subsystem.
These results will be demonstrated on a double- and a many-wall systems.
Next, I will show that closing the chain to form a ring gives rise to a family of periodic stripe 
solutions in the quiescent regime. These lattice structures are stable to small perturbations but 
large perturbation effect annihilations, which lead to transitions to other 
periodic solutions. It will be also argued that a cascade of such transitions may occur resulting in 
intriguing dynamics.

Consider an open-ended infinite chain prepared with an initial set of domain walls 
that occupy a finite region. The general analytic solution for $\psi$ can 
be represented in the form

\begin{equation}
\psi = \sum_{n=1}^N P_{N-n}(t) x^n = \prod_{n=1}^N (x - x_n(t))   
\label{eq:Ai} 
\end{equation}
Where $N$ is the total number of domain walls determined by the initial data. The coefficients 
$P_m$ are polynomials of order $m$ in $t$ whose structure has been discussed in reference \cite{BB}.
Real roots, $x_n$ correspond to physical domain walls in the field $\boeta$ where the 
angle $\theta$ goes through an abrupt change of $\pi$. 
The aim in the following is to find and analyse the dynamics of these walls.

First note that the field $\psi$ must satisfy Laplace's equation and therefore, carrying out 
the explicit derivatives, we get 

\begin{equation}
\nabla^2\psi = \sum_{n=1}^N \sum_{m\neq n}^N 
{{\dot x_n \dot x_m + 1}\over{(x - x_n)(x - x_m)}} - 
\sum_{n=1}^N {{\ddot x_n}\over{x - x_n}} = 0
\label{eq:Aiii} 
\end{equation}
where $x_n$ is the location of the $n$th wall.  
The solution for $\psi$ is valid for all $x$ and in particular when $x$ approaches the location of 
the $n$th wall, $x_n$. It follows that the residues of the poles at $x_n$ must all vanish. 
A straightforward calculation yields the equation of motion for the $n$th wall 

\begin{equation}
\ddot x_n - \sum_{m\neq n} {{1 + \dot x_n \dot x_m} \over {x_n - x_m}} = 0
\label{eq:Aiv} 
\end{equation}
This innocent-looking set of equations turns out to describe very rich dynamics and the rest of the 
paper is devoted to an analysis of its physical consequences.

There are several features to be noted about the set of equations (\ref{eq:Aiv}):
(1) The force on a wall is the sum of {\it two-body} forces between it and all the other walls 
along the chain. These forces are long ranged and depend on both position and velocities.
(2) Denoting the force on wall $n$ due to wall $m$ by $f_{nm}$ then $f_{nm} = -f_{mn}$, which 
is reminiscent of Newton's third law of action and reaction in translational invariant systems 
that this one is not.
(3) As a consequence of the above property, the system of walls conserves momentum. 
To see this define the centre of mass, $\gamma = \sum_{k=1}^N x_k/N$ and summ over $n$. 
The terms in the sum over $m$ cancel in pairs leading to $\ddot\gamma = 0$ and hence to 
$\dot\gamma = p_0 =$ constant. 
(4) Eqs (\ref{eq:Aiv}) are symmetric under time reversal $tt\to -t$. Therefore, for each  
forward solution $\psi_f(t)$ there exists a bacward one $\psi_b(t) = \psi_f(-t)$.

To illustrate the kinetics consider a double wall system. The equations of motion read 

\begin{equation}  
\ddot x_1 - {{\dot x_1 \dot x_2 + 1}\over{x_1 - x_2}} = 0
\label{eq:Avi} 
\end{equation} 
\begin{equation}
\ddot x_2 - {{\dot x_1 \dot x_2 + 1}\over{x_2 - x_1}} = 0
\label{eq:Avii} 
\end{equation} 
Defining the centre of mass velocity $\dot\gamma = (\dot x_1+\dot x_2)/2=p_0$ and their separation 
$\delta = (x_2 - x_1)/2$, one arrives at an equation for $\delta$

\begin{equation}
2\delta\ddot\delta + \dot\delta^2 - c^2 = 0
\label{eq:Aviii} 
\end{equation}
where $c = \sqrt{(1 + p_0^2)}$. This equation is symmetric not only under time reversal, but also under 
$\delta\to -\delta$. A particular solution of this equation is 

\begin{equation}
\delta_1 = \pm c t + \delta_0 
\label{eq:Aix}
\end{equation} 
where the positive (negative) sign corresponds to a solution that propagates forward (backward) in time.
To find the general solution define $\delta = c\ \eta^{2/3}$ and substitute into eq (\ref{eq:Aviii}), 
which gives

\begin{equation}
\ddot\eta - {3\over 4}\eta^{-1/3} = 0 
\label{eq:Bi}
\end{equation}
Integrating this equation we have

\begin{equation}
\dot\eta^2 - {9\over 4}\eta^{2/3} = E
\label{eq:Bia}
\end{equation}
with $E$ the integration constant. This equation can now be solved by quadratures 

\begin{equation}
t = t_0 \pm {{\chi}\over c} \left[\sqrt{{{\delta}\over{\chi}}\left(
1 + {{\delta}\over{\chi}}\right)} - {\text arcsinh}\left({{\delta}\over{\chi}}\right)^{1/2} \right] 
\label{eq:Bii} 
\end{equation}
where $\chi = 4cE/9$. Again the positive (negative) sign corresponds to forward (backward) propagation, 
which increases (decreases) monotonically with time. The case $E=0$ yields the linear solution (\ref{eq:Aix}). 
There can be three scenarios for the motion of the two walls: (a) moving apart, (b) approaching each other 
and eventually colliding, or (c) first approaching and then moving away. Which scenario takes place  
depends sensitively on initial conditions. Solutions (\ref{eq:Aix}) and (\ref{eq:Bii}) are 
real for all $E > 0$ when $\delta > 0$ and represent either the two walls moving apart forever 
(forward propagation) or moving into an eventual collision (backward propagation). In the latter case 
the time to collision is $t_{coll} = \delta_0 / c$ for the linear case and 
${{\chi}\over c} \left[\sqrt{{{\delta_0}\over{\chi}}\left(
1 + {{\delta_0}\over{\chi}}\right)} - {\text arcsinh}\sqrt{{{\delta_0}\over{\chi}}} \right]$
in the general case.
A careful analysis of $E < 0$ gives that in this case the walls first approach each other until 
$delta = \mid\chi\mid$ whereupon the solution switches to another branch and they move back out.
This happens when the approach is at sufficiently low velocities such that their interaction is 
always repelling. Once the walls have collided in the second scenario they annihilate. Whether they 
will reappear at a later time can be figured out from eq (\ref{eq:Ai}) for $\psi$.
It is worth noticing that on forward propagation the separation between the walls converges 
asymptotically to a steady increase at a rate of $\dot\delta = c$ that is independent of all 
initial conditions except the total momentum of the system $p_0$.   

Having understood the double wall case let us turn to many-wall systems.  
Consider two walls in such a system, say $n$ and $m$, whose velocities are, respectively,
$\dot x_n$ and $\dot x_m$. 
The interaction between these walls is repulsive when $\dot x_n \dot x_m > -1$ and is 
attractive otherwise. The attraction comes about when the two walls either move away from, 
or towards, each other at sufficiently high speeds. In the former case the attraction slows down 
too fast a separation, as is demonstrated in the double wall system where $\dot\delta\to c$. 
In the latter case, attraction between walls that approach each other too fast is bound to  
lead to a collision unless deflected by other walls in the system. On slow approaches walls
do not collide but rather achieve a minimal separation before departing in opposite 
directions.
The regime where all pairs satisfy $\dot x_n \dot x_m > -1$ is 'quiescent' in the sense 
that no walls can collide and annihilate. The number of walls in this regime remains constant 
and therefore so does the total momentum. Once the system 
exits the quiescent regime the conservation laws do not hold. Upon each annihilation event 
the number of walls changes and the system's momentum undergoes a  
discontinuous jump. The momentum lost in such a jump is exactly that of the 
two annihilating walls and it is the momentum of the remaining subsystem which remains 
conserved from the moment of annihilation onward.

\begin{figure}
\centerline{\psfig{file=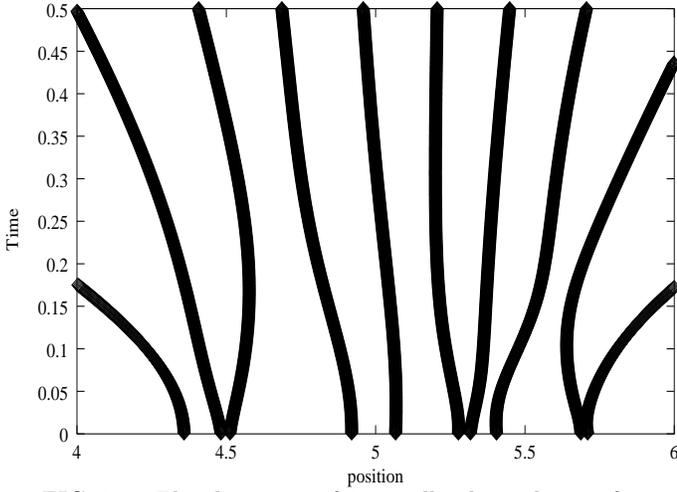,width=9.cm,height=6.5cm}}
\caption{
The dynamics of ten walls along the antiferromagnetic spin chain in the quiescent regime.
The walls initial positions are normally distributed around an arbitrary point along of the chain 
and they are assigned low velocities chosen from a Gaussian distribution of zero mean.
Note how the initial closure of the gap between some walls is reversed due to walls repulsion.}
\end{figure}

This analysis is supported by numerical studies where the above kinematic scenarios can be clearly 
observed in the dynamics of the walls. Figure 1 shows a system of ten
domain walls in the quiescent regime. The walls initial positions are randomly chosen from a Gaussian 
distribution around an arbitrary location along the chain. 
Their initial velocities are chosen from a Gaussian distribution of zero mean and a standard deviation
of 0.5 (arbitrary units). 
As expected, the walls fan out with time, demonstrating the repulsive interactions.
It can be observed that some neighbouring walls initially start approaching each other but their 
momentum is too low to effect collision and eventually they pull apart, as discussed above. 
Figure 2 shows those walls starting from the same initial positions but at initial velocities that 
are chosen from a Gaussian distribution whose standard deviation is three times as large. 
In this case some walls (eg, the third and fourth from the right) start off approaching at velocities 
that are sufficiently high to give rise to collision, upon which they collide and annihilate.
Computing the momentum of the system indeed gives that these are constant before and after annihilation 
at respective values of -0.39 and -1.72 (arbitrary units) with the change occuring abruptly  
at the moment of annihilation. 
 
\begin{figure}
\centerline{\psfig{file=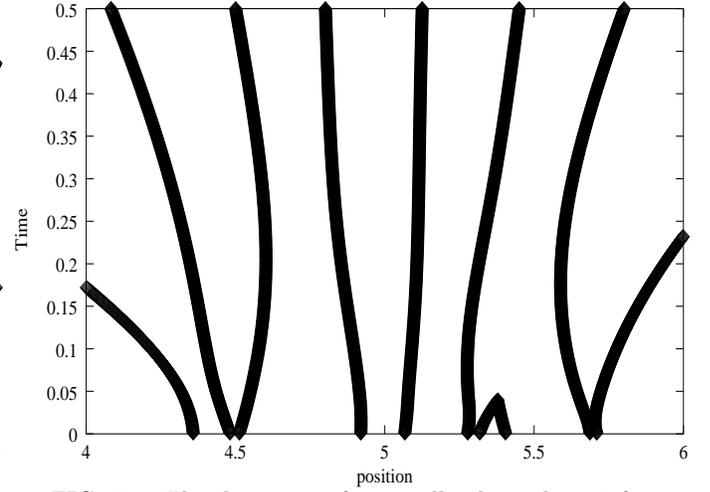,width=9.cm,height=6.5cm}}
\caption{
The dynamics of ten walls along the antiferromagnetic spin chain outside the quiescent regime.
The walls are initially positioned as in figure 1, but their random initial velocities are chosen from 
a Gaussian distribution that is three times as wide. 
Note the collision and annihilation of two neighbouring walls. Before (after) annihilation the total 
momentum of the system is constant at -0.39 (-1.72) with the change occuring abruptly at the 
moment of annihilation.}
\end{figure}

Let us turn now to finite chains that are closed to form rings. 
Examination of the equations of motion shows that they support periodic solutions of $N$ 
walls around the ring that are located at intervals of $a=L/N$, where $L$ is the chain length. 
These are stripe solutions that are stable in the quiescent regime. 
This can be seen by applying a perturbation to one of the walls in this state, moving it from its 
lattice position by a distance smaller than $a$ while keeping its angular velocity low enough 
so as not to exit the quiescent regime. 
The repulsive interaction with the closer neighbour dominates and pushes the wall back to its  
lattice position, as can be readily verified by linear perturbation analysis.  
This means that in the quiescent regime the stripe lattice dynamics consists only of oscillations 
around the periodic structure. The spectrum of these oscillations would be interesting to analyse 
but this is outside the thrust of this paper.

A most intriguing feature of these solutions is that rotating the entire stripe lattice rigidly 
at an arbitrary angular velocity $\omega$ is also a solution. Moreover, 
the amplitude of the repulsive forces between the walls increases with the angular velocity as 
$ 1 + (\omega L/2\pi)^2$. 
The rotating solution is again stable under a perturbation of the walls from their lattice positions. 
However, the stability of the periodic structure depends more on perturbations to their angular 
velocities then positions. Suppose we give wall $n$ an excessive angular velocity $\delta\omega$. 
If $\delta\omega$ is in the same direction as $\omega$ this wall experiences a net restoring force 
from its two neighbours. The same applies if $\delta\omega$ is in the opposite direction to 
$\omega$ but its magnitude is smaller than $\omega\left[ 1 + \left( 2\pi / L\omega \right)^2 \right]$. 
Being stable, the dynamics of the lattice in the rotating frame due to low kinetic energy excitations 
will again consist of oscillations of the walls around the lattice positions. 
Once $\delta\omega$ exceeds the critical value of $\omega\left[ 1 + \left( 2\pi / L\omega \right)^2 \right]$ 
in the $ - \omega$ direction, the net force on the $n$th wall becomes attractive and pulls it 
further in that direction. The $n$th wall then runs into the $(n-1)$th wall and they annihilate. 
Thus, too high excessive angular velocities to individual walls destabilises the lattice. 

The dynamics ensuing in the wake of such an instability are fascinating:
Upon annihilation a gap appears in the periodic structure and the remaining $N-2$ walls must shift 
to accomodate the new conditions. If this readjustment gives rise only to low excessive angular velocities 
then the system settles, without further ado, into a new structure of period $a = L/(N-2)$. 
If, however, the readjustment gives rise to high excessive velocities in the $-\omega$ direction 
then another annihilation may occur, leading to further damage of the lattice. 
Thus, a cascade of annihilation events may follow the initial instability leading to strange and 
rich dynamics. 
This author believes that such behaviour can be observed by monitoring directly the spectrum of 
the stripe lattice to vibrational excitations.
At low kinetic energies \cite{energy} one expects plasmon-like excitations due to the long range Coulomb 
interaction. But, at higher kinetic energies the spectrum should become muddled by the above dynamics.

To conclude, the dynamics of domain walls on classical antiferromagnetic Heisenberg chain have been found to 
be surprisingly rich. The equation of motion of the walls along the chain has been derived from first 
principles, showing that the walls experience only two-body forces. 
These forces consist of both a Coulomb-like term and a long range velocity dependent term. 
The interactions can be either repulsive, in a quiescent regime, or attractive at high velocities.
The system of walls has been found to conserve momentum as long as the number of walls is conserved, which 
is always the case in the quiescent regime. Outside this regime walls may collide and annihilate, upon 
which the total momentum of the system changes discontinuously to a new level that remain constant until 
another annihilation event occurs.
An explicit solution of a double wall system has been presented, giving insight into the dynamics of 
many-wall systems.
A numerical solution of the latter has been discussed both inside and outside the quiescent regime and 
annihilation events have been demonstrated with increasing wall velocities.  
This solution also demonstrated the above principle of momentum conservation between annihilations.

On circular chains the equations of motion give rise to periodic stripe solutions that are stable for 
small perturbations. These periodic structures can roitate as a rigid body with the structures 
becoming {\it more stable} with increasing angular velocity. This observation may have direct application
in rotating magnetic memory devices. 
The dynamics of the stripe lattice has been found to be very rich. In the quiescent regime it  
oscillates and at low kinetic energies the system should resemble a confined electron gas due to the 
long range Coulomb-like repulsion. Increasing the kinetc energy brings into play the nonlinearity of 
the equation of motion distorting the oscillation spectrum into potentially new and interesting 
configurations. Nevertheless, the dynamics should still be confined to oscilatory vibrational 
excitations. 
As the kinetc energies increase further the system exits the quiescent regime, interactions turn 
attractive and walls can collide and annihilate. 
Annihilations on a ring lead to rearrangements of the walls, which in turn, can either settle into a new 
periodic structure or result in a cascade of further annihilation events, the dynamics of which 
should be very interesting to study.


\begin{thebibliography}{99}
\bibitem{energy} Although energy is not conserved in the system, the kinetic energy is nevertheless a 
useful parameter to characterise the quiescent regime
\end{thebibliography}
\end{document}